\documentclass[aps,prl,showpacs,twocolumn,floats]{revtex4}

\usepackage{graphicx,psfig}% Include figure files
\usepackage{dcolumn}% Align table columns on decimal point
\usepackage{bm}% bold math

\begin{document}
\title{Surface contribution to the anisotropy of magnetic nanoparticles \vspace{-1mm%
} }
\author{D. A. Garanin$^{1}$ and H. Kachkachi$^{2}$}
\affiliation{$^{1}$Institut f\"{u}r Physik, Johannes Gutenberg-Universit\"{a}t, D-55099 Mainz, Germany\\
$^{2}$Laboratoire de Magn\'{e}tisme et d'Optique, Universit\'{e}
de Versailles St. Quentin, 45 av. des Etats-Unis, 78035
Versailles, France}

\begin{abstract} We calculate the contribution of the N\'{e}el
surface anisotropy to the effective anisotropy of magnetic
nanoparticles of spherical shape cut out of a simple cubic
lattice. The effective anisotropy arises because deviations of
atomic magnetizations from collinearity and thus the energy
depends on the orientation of the global magnetization. The result
is second order in the N\'{e}el surface anisotropy, scales with
the particle's volume and has cubic symmetry with preferred
directions $[\pm 1,\pm 1,\pm 1].$
\end{abstract}

 \maketitle

With the decreasing size of magnetic particles, surface effects
are believed to become more and more pronounced. A simple argument
based on the estimation of the fraction of surface atoms shows
that for a particle of spherical shape and diameter $D$ (in units
of the lattice spacing), this fraction is an appreciable number of
order $6/D$. Regarding the fundamental property of magnetic
particles, the magnetic anisotropy, the role of surface atoms is
augmented by the fact that these atoms in many cases experience
surface anisotropy (SA) that by far exceeds the bulk anisotropy.
As was suggested by N\'{e}el \cite{nee54} and microscopically
shown in Ref.\ \cite{vicmac93}, the leading contribution to the
anisotropy is due to pairs of atoms and can be written as
\begin{equation}
{\cal H}_{A}=\frac{1}{2}\sum_{ij}L_{ij}\left( {\bf m}_{i}\cdot {\bf e}%
_{ij}\right) ^{2}+\ldots ,\qquad \text{ \mbox{$\vert$} }{\bf
m}_{i}|=1,  \label{NSADef}
\end{equation}
where ${\bf m}_{i}$ is the reduced magnetization (spin
polarization) of the $i$th atom, ${\bf e}_{ij}$ are unit vectors
directed from the $i$th atom to its neighbors, and $L_{ij}$ is the
pair-anisotropy coupling that depends on the distance between
atoms. Eq.\ (\ref{NSADef}) describes in a unique form both the
bulk anisotropy including the effect of elastic strains and the
effect of missing neighbors at the surface that leads to the SA.
In particular, for an unstrained simple cubic (sc) lattice the
bulk anisotropy in Eq.\ (\ref {NSADef}) disappears since
$m_{x}^{2}+m_{y}^{2}+m_{z}^{2}=1$ is an irrelevant constant, and
one has to take into account the dropped (much smaller) terms of
Eq.\ (\ref{NSADef}) that yield the cubic bulk anisotropy. On the
other hand, surface atoms experience (large) anisotropy of order
$L$ due
to the broken symmetry of their crystal environment -- the so-called N\'{e}%
el surface anisotropy (NSA). These atoms can make a contribution
to the {\em effective} volume anisotropy decreasing as $1/D$ with
the particle's linear size: $K_{V,{\rm eff}}=K_{V}+K_{S}/D,$ as
was observed in a number of experiments (see, e.g., Refs.\
\cite{resetal98,chekitokashi99}).

\begin{figure}[t]
\unitlength1cm
\begin{picture}(11,6)
\centerline{\psfig{file=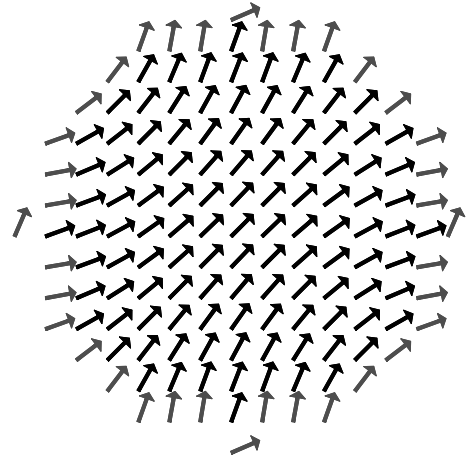,angle=-90,width=6cm}}
\end{picture}
\caption{ \label{nsa-structure} Magnetic structure of a spherical
nanoparticle of linear size $N=15$ with $L/J=2$ for the global
magnetization directed along [1,1,0], showing atoms in the plane
$z=0$.}
\end{figure}%

The $1/D$ surface contribution to $K_{V,{\rm eff}}$ is in accord
with the picture of all magnetic atoms tightly bound by the
exchange interaction whereas only the surface atoms feel the
surface anisotropy. This is definitely true for magnetic films
where a huge surface contribution to the effective anisotropy has
been observed. The same is the case for cobalt nanoclusters of the
form of truncated octahedrons \cite{jametal01} where contributions
from different faces, edges, and apexes compete resulting in a
nonzero, although significantly reduced, surface contribution to
$K_{V,{\rm eff}}$. However, for symmetric particle shapes such as
cubes or spheres, the symmetry leads to vanishing of this
(first-order) contribution. In this case one has to take into
account deviations from the collinearity of atomic spins that
result from the competition of the SA and the exchange interaction
$J$. The resulting structures (for the simplified radial SA model)
can be found in Refs.\ \cite
{dimwys9494Cobalt,labetal02,dimkac0202prb} (see also Fig.\ \ref
{nsa-structure} for the NSA). In the case $L\gtrsim J$ deviations
from collinearity are very strong, and it is difficult if not
impossible to characterize the particle by a global magnetization
suitable for the definition of the effective anisotropy. On the
other hand, in the typical case $L\ll J$ \ the magnetic structure
is nearly collinear with small deviations that can be computed
perturbatively in $L/J\ll 1$. The global magnetization vector
${\bf m}_{0}$ can be used to define the anisotropic energy of the
whole particle. The key point is that deviations from
collinearity and thus the energies of the system are different for
different orientations of ${\bf m}_{0},$ even for a particle of a
spherical shape, due to  the crystal lattice. For the latter the
overall anisotropy per unit cell is proportional to $L^{2}/J,$
i.e., it scales with the particle's volume.

The aim of this Letter is to illustrate this idea by calculating
the second-order contribution from the NSA to the effective
particle's anisotropy for the minimal model of a magnetic
nanoparticle of spherical shape cut out of a sc lattice. We will
neglect the small cubic anisotropy and magnetostatic effects for
the sake of transparency. The problem will be solved numerically
on the lattice by minimizing the energy with the help of a damped
Landau-Lifshitz equation without the precession term, with the
average particle's magnetization constrained in a desired
direction. We also produce an analytical solution in the
continuous limit of larger particles that will be shown to agree
with the numerical solution.

We consider the nearest-neighbor form of Eq.\ (\ref{NSADef}) with
the unique constant $L.$ For a sc lattice it reduces to
\begin{equation}
{\cal H}_{A}=\sum_{i}{\cal H}_{Ai},\qquad {\cal H}_{Ai}=\frac{L}{2}%
\sum_{\alpha =x,y,z}z_{i\alpha }m_{i\alpha }^{2},  \label{NSAsc}
\end{equation}
where $z_{i\alpha }=0,1,2$ are the numbers of available nearest neighbors of
the atom $i$ along the axis $\alpha .$ One can see that the NSA is in
general biaxial. For $L>0$ and $z_{i\alpha }=0<z_{i\beta }=1<z_{i\gamma }=2$
the $\alpha $-axis is the easy axis and the $\gamma $-axis is the hard axis.
If the local magnetizations ${\bf m}_{i}$ are all directed along one of
the crystallographic axes $\alpha ,$ then the anisotropy fields ${\bf H}%
_{Ai}=-\partial {\cal H}_{Ai}/\partial {\bf m}_{i}$ are also directed along $%
\alpha $ and are thus collinear with ${\bf m}_{i}$. Hence, at
least for $L\ll J,$ there are no deviations from collinearity if
the global magnetization ${\bf m}_{0}$ is directed along one of
the crystallographic axes. For other orientations of ${\bf
m}_{0}$, the vectors $ {\bf m}_{i}$ and ${\bf H}_{Ai}$ are not
collinear, and the transverse component of ${\bf H}_{Ai}$ with
respect to ${\bf m}_{i}$ causes a slight canting of ${\bf m}_{i}$
and thereby a deviation from the collinearity of magnetizations on
different sites. This adjustment of the magnetization to the
surface anisotropy leads to the lowering of energy. As we shall
see, this effect is strongest for the [$\pm $1,$\pm $1,$\pm $1]
orientations of ${\bf m}_{0}$. For both signs of $L$ these are
easy orientations, whereas [$\pm 1,0,0],$ [$0,\pm 1,0],$ and
[$0,0,\pm 1]$  are hard orientations.

\begin{figure}[t]
\unitlength1cm
\begin{picture}(11,6)
\centerline{\psfig{file=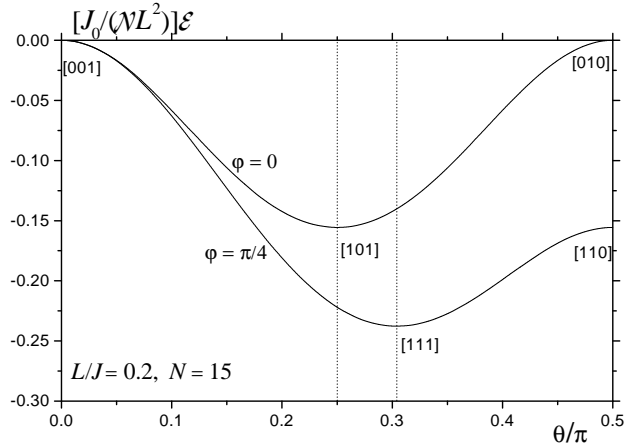,angle=-90,width=9cm}}
\end{picture}
\caption{ \label{nsa-etheta} Reduced shifted energy of the
particle for different orientations of its global magnetization,
obtained from Eq.\ (\protect\ref{LLEqs}). These curves manifest
the cubic symmetry of the effective anisotropy, see Eq.\
(\protect\ref{Ekappa}).}
\end{figure}

We consider here explicitly spherical particles cut out of a cube
with dimensions $N\times N\times N$ in the units of the atomic
spacing. If an atom is within or exactly on the sphere with the diameter $%
D=N-1,$ it belongs to the particle. The number of atoms in the particle $%
{\cal N}$ approaches ${\cal N}\cong (\pi /6)(N-1)^{3}$ for
$N\gtrsim 10,$ with fluctuations for smaller $N.$ Our numerical
results for the magnetic energy of spherical particles as a
function of the orientation of the global (average) magnetization
are shown in Fig.\ \ref{nsa-etheta}. They confirm the statements
of the previous paragraph.

To produce Fig.\ \ref{nsa-etheta}, we use the classical
Hamiltonian
\begin{equation}
{\cal H}=-\frac{1}{2}\sum_{ij}J_{ij}{\bf m}_{i}\cdot {\bf m}_{j}+{\cal H}_{A}
\label{HamDef}
\end{equation}
with the nearest-neighbor exchange coupling $J$ and ${\cal H}_{A}$ of Eq.\ (\ref{NSAsc}%
). To fix the global magnetization of the particle in a desired
direction ${\bm \nu }_{0}$ (${\bm \nu }_{0}|=1$), we use the
energy function with a Lagrange multiplier ${\bm \lambda }$:
\begin{equation}
{\cal F}={\cal H}-{\cal N}{\bm \lambda \cdot }\left( {\bm \nu }-{\bm \nu }%
_{0}\right) ,\qquad {\bm \nu \equiv }\frac{\sum_{i}{\bf
m}_{i}}{\left| \sum_{i}{\bf m}_{i}\right| }.  \label{FFuncDef}
\end{equation}
To minimize ${\cal F},$ we solve the evolution equations
\begin{eqnarray}
{\bf \dot{m}}_{i} &=&-\left[ {\bf m}_{i}\times \left[ {\bf m}_{i}\times {\bf %
F}_{i}\right] \right] ,\qquad {\bf F}_{i}\equiv -\partial {\cal F}/\partial
{\bf m}_{i}  \nonumber \\
{\dot{\bm \lambda}} &=&{\bf \partial }{\cal F}/\partial {\bm \lambda =-}%
{\cal N}\left( {\bm \nu }-{\bm \nu }_{0}\right),  \label{LLEqs}
\end{eqnarray}
starting from ${\bf m}_{i}={\bm \nu }_{0}={\bf m}_{0}$ and ${\bm
\lambda =0,} $ until a stationary state is reached. In this state
${\bm \nu }={\bm \nu }_{0}$ and $\left[ {\bf m}_{i}\times {\bf
F}_{i}\right] =0,$ i.e., the torque due to the term ${\cal N}{\bm
\lambda \cdot }\left( {\bm \nu }-{\bm \nu }_{0}\right) $ in ${\cal
F}$ compensates for the torque acting to rotate the
global magnetizations towards the minimum-energy directions [$\pm $1,$\pm $1,%
$\pm $1]. Since the former torque is unphysical, this method is
applicable only for a small surface anisotropy, so that both
torques are small, and adding a small formal compensative torque
does not strongly distort the magnetic structure.

In physical terms, the existence of the well-defined state with a
given orientation of the global magnetization can be justified as
follows. For $ L\ll J,$ the relaxation of the magnetization splits
into two stages. The first stage, adjustment of the magnetic
structure to the surface anisotropy, involves energies of order
$L$ and is relatively fast. The second stage, rotation of the
global magnetization to the global energy minimum with the
magnetic structure adjusted at any moment, involves energies of
order $ L^{2}/J$ and is much slower. Introducing the
global-orientation constraint above eliminates the second stage of
the relaxation, so that the result of the first relaxation stage
is seen in pure form.

\begin{figure}[t]
\unitlength1cm
\begin{picture}(11,6)
\centerline{\psfig{file=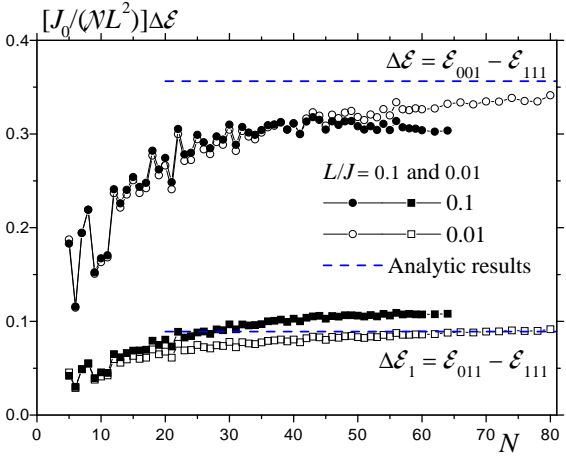,angle=-90,width=9cm}}
\end{picture}
\caption{ \label{nsa-en} Differences of the particle energies
between main orientations of the global magnetization vs the
particle size in the scaled form for $L/J=0.1$ and 0.01. The
scaling is valid for $N\lesssim J/L$, and its violation for
$L/J=0.1$ is seen in the right part of the figure. }
\end{figure}%

Fig.\ \ref{nsa-en} shows the dependence of the normalized particle
energy differences between the basic directions [001], [011], and
[111]. One can see that $\Delta E/{\cal N}$ tends to a large-$N$
limit, i.e., for large linear sizes $N$ the energy differences due
to the SA scale with particle's volume $V\propto {\cal N}\sim
(N-1)^{3}$. These results suggest that the problem can be solved
analytically with the help of the continuous approximation for
$N\gg 1.$ To this end, we replace in Eq.\ (\ref{NSAsc}) the number
of nearest neighbors of a surface atom by its average value
\begin{equation}
z_{i\alpha }\Rightarrow \overline{z}_{i\alpha }=2-|n_{\alpha }|/\max \left\{
|n_{x}|,|n_{y}|,|n_{z}|\right\} .  \label{zAvr}
\end{equation}
Here $n_{\alpha }$ is the $\alpha $-component of the normal to the surface $%
{\bf n.}$ The surface-energy density can then be obtained by
dropping the constant term and
multiplying Eq.\ (\ref{NSAsc}) by the surface  atomic density $f({\bf n}%
)=\max \left\{ |n_{x}|,|n_{y}|,|n_{z}|\right\}$:
\begin{equation}
E_{S}({\bf m,n})=-\frac{L}{2}\left[
|n_{x}|m_{x}^{2}+|n_{y}|m_{y}^{2}+\left| n_{z}\right|
m_{z}^{2}\right] .  \label{ESDef}
\end{equation}
At equilibrium, in the continuous approximation the Landau-Lifshtz
equation reduces to
\begin{equation}
{\bf m\times H}_{{\rm eff}}=0,\qquad {\bf H}_{{\rm eff}}={\bf H}_{A}+J\Delta
{\bf m,}  \label{LLEqEqui}
\end{equation}
where $\Delta $ is the Laplace operator and the anisotropy field
\begin{equation}
{\bf H}_{A}=-\frac{dE_{S}}{d{\bf m}}\delta (r-R),\qquad R\equiv \frac{1}{2}%
\left( N-1\right) .  \label{HADelta}
\end{equation}
For $L\ll J$ the deviations of ${\bf m(r)}$ from the homogeneous state $%
{\bf m}_{0}$ are small and one can linearize the problem:
\begin{equation}
{\bf m(r)\cong m}_{0}+{\bm \psi }({\bf r,m}_{0}),\qquad \psi
\equiv |{\bm \psi }|\ll 1.  \label{mAnsatz}
\end{equation}
The correction ${\bm \psi }$ is the solution of the internal
Neumann boundary problem for a sphere
\begin{eqnarray}
&&\Delta {\bm \psi }=0,\qquad \left. \frac{\partial {\bm \psi }}{%
\partial r}\right| _{r=R}={\bf f(m,n)}  \nonumber \\
&&{\bf f}=-\frac{1}{J}\left[ \frac{dE_{S}({\bf m,n})}{d{\bf m}}-\left( \frac{%
dE_{S}({\bf m,n})}{d{\bf m}}\cdot {\bf m}\right) {\bf m}\right] ,
\label{psiNeumann}
\end{eqnarray}
where ${\bf n\equiv r/}R$ and ${\bf m}$ stands for ${\bf m}_{0}$
with the index 0 dropped for transparency. $\psi$ has the form
\begin{equation}
{\bm \psi }({\bf r,m})=\frac{1}{4\pi }\int_{S}d^{2}{\bf r}^{\prime }G({\bf r,r}%
^{\prime }){\bf f(m,n}^{\prime }{\bf )}  \label{psiSol}
\end{equation}
with the Green function
\begin{eqnarray}
G({\bf r,r}^{\prime }) &=&\frac{1}{\left| {\bf r-r}^{\prime }\right| }+\frac{%
R}{S({\bf r,r}^{\prime })}+\frac{1}{R}\ln \frac{R^{2}}{R^{2}-{\bf r\cdot r}%
^{\prime }+S({\bf r,r}^{\prime })}  \nonumber \\
S({\bf r,r}^{\prime }) &\equiv &\sqrt{R^{4}+{\bf r}^{2}{\bf
r}^{\prime 2}-2R^{2}({\bf r\cdot r}^{\prime })}.  \label{GDef}
\end{eqnarray}
One can make the estimation
\begin{equation}
 \psi({\bf r,m})\sim RL/J\sim NL/J,\qquad |{\bf r|}=R.
\label{psiEstim}
\end{equation}
This shows that for whatever small values of $L$ the applicability
condition of our linearization method $\psi \equiv |{\bm \psi
}|\ll 1$ will be invalidated for sufficiently large particle
sizes.

Now we are prepared to calculate the magnetic energy of the nanoparticle.
Dropping the trivial constant term leads to the second-order energy
\begin{equation}
{\cal E}_{2}\cong {\cal E}_{2,V}+{\cal E}_{2,S}=\int_{V}d^{3}{\bf r}\frac{J}{%
2}\left( \nabla {\bm \psi }\right) ^{2}+\int_{S}d^{2}{\bf r}\left( \frac{%
dE_{S}}{d{\bf m}}\cdot {\bm \psi }\right)  \label{EVS}
\end{equation}
that is a sum of the inhomogeneous exchange and anisotropy
energies. With the help of Eq.\ (\ref{psiNeumann}) this yields
\begin{equation}
{\cal E}_{2}\cong \frac{1}{4\pi }\frac{1}{2J}\int \int_{S}d^{2}{\bf r}d^{2}%
{\bf r}^{\prime }G({\bf r,r}^{\prime })\Phi ({\bf m},{\bf n,n}^{\prime })
\label{ERes}
\end{equation}
with
\begin{eqnarray}
\Phi ({\bf m},{\bf n,n}^{\prime }) &=&\left( {\bf m}\cdot \frac{dE_{S}({\bf %
m,n})}{d{\bf m}}\right) \left( {\bf m}\cdot \frac{dE_{S}({\bf m,n}^{\prime })%
}{d{\bf m}}\right)  \nonumber \\
&&-\left( \frac{dE_{S}({\bf m,n})}{d{\bf m}}\cdot \frac{dE_{S}({\bf m,n}%
^{\prime })}{d{\bf m}}\right) .  \label{PhiDef}
\end{eqnarray}
The first term in $\Phi ({\bf m},{\bf n,n}^{\prime })$ can be simplified
using $ {\bf m}\cdot dE_{S}({\bf m,n})/d{\bf m} =2E_{S}({\bf m,n%
})$ following from Eq.\ (\ref{ESDef}). The second term in $\Phi ({\bf m},{\bf %
n,n}^{\prime })$ is quadratic in the magnetization components and
contributes only with the irrelevant term proportional to
$m_{x}^{2}+m_{y}^{2}+m_{z}^{2}=1$ to the energy. Thus ${\cal E}_2$
simplifies to
\begin{equation}
{\cal E}_{2}\cong \frac{1}{2\pi J}\int \int_{S}d^{2}{\bf r}d^{2}{\bf r}%
^{\prime }G({\bf r,r}^{\prime })E_{S}({\bf m,n})E_{S}({\bf
m,n}^{\prime }), \label{EFinal}
\end{equation}
that is of fourth order in the global-magnetization components $%
m_{\alpha }.$ Taking into account the cubic symmetry and computing
numerically a double surface integral one can write the result of
Eq.\ (\ref {EFinal})\ as
\begin{equation}
{\cal E}_{2}\cong \kappa \frac{L^{2}{\cal N}}{J_{0}}\left(
m_{x}^{4}+m_{y}^{4}+m_{z}^{4}\right) ,\qquad \kappa =0.53465,  \label{Ekappa}
\end{equation}
where $J_{0}=zJ=6J.$ This defines the large-$N$ asymptotes in
Fig.\ \ref{nsa-en} that are shown by the horizontal lines.

The analytical results above are valid for particle sizes $N$ in
the range
\begin{equation}
1\ll N\ll J/L.  \label{NApplReg}
\end{equation}
The lower boundary is the applicability condition of the
continuous approximation. Since the surface of a nanoparticle is
made of atomic terraces separated by atomic steps, each terrace
and each step with its own form of NSA [see Eq.\ (\ref{NSAsc})],
the variation of the local NSA along the surface is very strong.
Approximating this variation by a continuous function according to
Eq.\ (\ref{zAvr}) requires pretty large particle sizes $N.$ This
is manifested by a slow convergence to the large-$N$ results in
Fig.\ \ref{nsa-en}.

The upper boundary in Eq.\ (\ref{NApplReg}) is the applicability
condition of the linear approximation in ${\bf \psi },$ see Eqs.\
(\ref{mAnsatz}) and (\ref {psiEstim}). For $N\gtrsim J/L$
deviations from the collinear state are strong, and the effective
anisotropy of a magnetic nanoparticle cannot be introduced. The
solution found above becomes invalid even for orientations of the
global magnetization along the crystallographic axes where ${\bm
\psi=0}$. In this case those surface spins close to the equatorial
plane (${\bf n} \bot {\bf m}$) for $L>0$ or to the poles (${\bf
n}\Vert {\bf m}$) for $L<0$ develop instability and turn away from
${\bf m}$ for $N\gtrsim J/L.$ Gradual disappearance of the
collinear magnetic structure of a particle with increasing size
stems from the ``softnening'' of the exchange interaction at large
distances. A related phenomenon is the breakdown of the
single-domain state of particles with a uniaxial bulk anisotropy
with increasing size due to the magnetostatic effect.

As we have seen in Eq.\ (\ref{Ekappa}), the contribution of the SA
into the
overall anisotropy of a magnetic particle scales with its volume $%
V\propto N^{3}\sim {\cal N}.$ This surprising result, that
contradicts the initial guess on the role of the surface effects
based on the ratio of the numbers of surface and volume spins
$\sim 6/D$, is due to the penetration of perturbations from the
surface deeply into the bulk. If a uniaxial bulk anisotropy
$D_{V}$ is present in the system, perturbations from the surface
will be screened at the bulk correlation length (or the domain-wall width) $%
\delta \sim \sqrt{J/D_{V}}.$ Then for $D\sim N\gtrsim \delta $ the
contribution of the SA to the overall anisotropy will scale as the surface: $%
{\cal E}_{2}\sim \left( L^{2}/J\right) N^{2}\delta .$ As follows from Eq.\ (%
\ref{NApplReg}), this regime requires $D_{V}\gtrsim L^{2}/J,$ i.e., the
dominance of the bulk anisotropy over the SA in the overall anisotropy.

In most cases the bulk anisotropy is much smaller than the surface
anisotropy for the microscopic reasons discussed at the beginning
of this Letter. Then, at least for not too large particles,
$N\lesssim \delta $, contributions of both anisotropies to the
overall anisotropy are additive and scale as the volume. If the
bulk anisotropy is cubic, both contributions have the same cubic
symmetry [see Eq.\ (\ref{Ekappa})], and the experiment should
yield a value of the effective cubic anisotropy different form the
bulk value\cite{jametal01}.
For the uniaxial bulk anisotropy, the two contributions have
different functional forms. Even if the bulk anisotropy is
dominant so that the energy minima are realized for ${\bf m}\Vert
{\bf e}_{z},$ the surface anisotropy makes the energy dependent on
the azimuthal angle $\varphi.$ This changes the type of the energy
barrier for the particle creating saddle points. The latter, in
particular, strongly influences the process of thermal activation
of magnetic particles \cite{garkencrocof99}.

We stress that we have calculated the second-order contribution of
the N\'eel surface anisotropy to the effective anisotropy of a
magnetic particle, and this is the only effect for symmetric
particle shapes such as cubic or spherical. For small deviations
from this symmetry, i.e., for weakly elliptic or weakly
rectangular particles, there is a correspondingly weak first-order
contribution ${\cal E}_{1}$ that adds up with our
second-order contribution. For an ellipsoid with axes $a$ and $%
b=a(1+\epsilon )$, $\epsilon \ll 1,$ one has ${\cal E}_{1}\sim L%
{\cal N}^{2/3}\epsilon m_{z}^{2}$ [cf. Eq.\ (\ref{Ekappa})], so
that
\begin{equation}
\frac{{\cal E}_{2}}{{\cal E}_{1}}\sim \frac{L}{J}\frac{N}{\epsilon }
\label{E2E1Ratio}
\end{equation}
can be large even for $L/J\ll 1.$ Whereas ${\cal E}_{1}$ scales
with the particle's surface and can be experimentally identified
as a surface contribution, ${\cal E}_{2}$ scales with the volume
and thus renormalizes the volume anisotropy of nanoparticles.

The N\'{e}el constant $L$ is in most cases poorly known. However,
for metallic Co
Ref.\ \cite{chubalhan94} quotes the value of SA $-1.5\times 10^{8}$ erg/cm$%
^{3},$ i.e., $L\sim -10$ K. This is much smaller than $J\sim
10^{3}$ K, which makes our theory valid for particle sizes up to
$N\sim J/L\sim 100,$
according to Eq.\ (\ref{NApplReg}). For this limiting size one has ${\cal E}%
_{2}/{\cal E}_{1}\sim 1/\epsilon $ that is large for nearly spherical
particles, $\epsilon \ll 1.$

We are indebted to R. Schilling for critical reading of the
manuscript. D. G. thanks A. A. Lokshin for a valuable discussion.

%\vspace{-0.5cm}
\bibliographystyle{prsty}
%\bibliography{gar-oldworks,gar-surface-nano,gar-own}

\end{document}